\documentclass[11pt]{article}

\usepackage{amssymb}
\usepackage[numbers,sort&compress]{natbib}
\usepackage{amsmath,amsfonts,graphicx,epsfig}
\usepackage{bm}
\usepackage{color}

\def\bea{\begin{eqnarray}}
\def\eea{\end{eqnarray}}
\def\be{\begin{equation}}
\def\ee{\end{equation}}
\def\ba{\begin{array}}
\def\ea{\end{array}}
\def\d{\partial}

\def\nn{\nonumber}

\newcommand{\red}[1]{{\color{black} #1}}

\begin{document}

\setlength\arraycolsep{2pt}

\renewcommand{\theequation}{\arabic{section}.\arabic{equation}}
\setcounter{page}{1}

\begin{titlepage}

\begin{center}

\vskip 1.5 cm

{\huge \bf A generalized non-Gaussian consistency relation for single field inflation}

\vskip 2.0cm

{\Large 
Rafael Bravo$^{a}$, Sander Mooij$^{a,b}$, Gonzalo A. Palma$^{a}$\\ and Basti\'an Pradenas$^{a}$
}

\vskip 0.5cm

{\it $^{a}$Grupo de Cosmolog\'ia y Astrof\'isica Te\'orica, Departamento de F\'{i}sica, FCFM, \mbox{Universidad de Chile}, Blanco Encalada 2008, Santiago, Chile. \\
$^{b}$Institute of Physics, Laboratory of Particle Physics and Cosmology, \'Ecole Polytechnique F\'ed\'erale de Lausanne, CH-1015 Lausanne, Switzerland.  }

\vskip 2.5cm

\end{center}

\begin{abstract} 

We show that a perturbed inflationary spacetime, driven by a canonical single scalar field, is invariant under a special class of coordinate transformations together with a field reparametrization of the curvature perturbation in co-moving gauge. This transformation may be used to derive the squeezed limit of the 3-point correlation function of the co-moving curvature perturbations valid in the case that these do not freeze after horizon crossing. This leads to a generalized version of Maldacena's non-Gaussian consistency relation in the sense that the bispectrum squeezed limit is completely determined by spacetime diffeomorphisms. Just as in the case of the standard consistency relation, this result may be understood as the consequence of how long-wavelength modes modulate those of shorter wavelengths. This relation allows one to derive the well known violation to the consistency relation encountered in ultra slow-roll, where curvature perturbations grow exponentially after horizon crossing. 

\end{abstract}

\end{titlepage}

\newpage

\setcounter{equation}{0}
\section{Introduction}

It is by now well understood that Maldacena's consistency relation $f_{\rm NL} = 5 (1 - n_s)/12$~\cite{Maldacena:2002vr}, linking together the amount of local (squeezed) non-Gaussianity $f_{\rm NL}$ with the spectral index $n_s-1$, and valid for attractor models of single field inflation~\cite{Creminelli:2004yq, Seery:2005wm, Chen:2006nt, Cheung:2007sv, Ganc:2010ff, RenauxPetel:2010ty,Kundu:2014gxa, Kundu:2015xta,Gong:2017yih}, cannot be directly observed. A correct account of the observable amount of primordial local non-Gaussianity yields~\cite{Tanaka:2011aj, Pajer:2013ana, Dai:2015rda, Cabass:2016cgp, Tada:2016pmk}
\be
f^{\rm obs}_{\rm NL} =  0 + \mathcal O (k_L / k_S)^2,  \label{fNL=0}
\ee
where $\mathcal O (k_L / k_S)^2$ stands for non-Gaussianity produced by non-primordial phenomena such as gravitational lensing and redshift perturbations (the so called projection effects~\cite{Baldauf:2011bh, Yoo:2009au}). This result may be understood as coming from a cancellation between the primordial value predicted in co-moving gauge $5 (1 - n_s)/12$, and a correction $-5 (1 - n_s)/12 + \mathcal O (k_L / k_S)^2$ that arises after considering a change of coordinates rendering gauge invariant observables. This coordinate change corresponds to a transformation from co-moving coordinates to the so called conformal Fermi coordinates~\cite{Manasse:1963zz, Pajer:2013ana}. 

It appears to be entirely reasonable that the cancellation leading to (\ref{fNL=0}) is only effective when the prediction of primordial non-Gaussianity corresponds to $f_{\rm NL} = 5 (1 - n_s)/12$. This is because Maldacena's consistency relation itself may be thought of as the consequence of a space-time reparametrization linking short- and long-wavelength curvature perturbations re\-a\-li\-zed with the help of a symmetry of the system under a simultaneous spatial dilation and a field reparametrization~\cite{Creminelli:2004yq}. Thus, any measurement of local non-Gaussianity would directly rule out single field models of slow-roll inflation~\cite{Guth:1980zm, Linde:1981mu, Starobinsky:1980te, Albrecht:1982wi, Mukhanov:1981xt} (attractor models of inflation), but it would not rule out other classes of inflation. In particular, one would be seriously motivated to consider more exotic models of inflation such as curvaton scenarios~\cite{Sasaki:2006kq}, multi-field models~\cite{Byrnes:2008wi}, or non-attractor models of inflation (that is, models for which the background depends on the initial conditions~\cite{Kinney:2005vj, Namjoo:2012aa, Martin:2012pe, Chen:2013aj, Mooij:2015yka}). For instance, in the case of ultra slow-roll inflation~\cite{Tsamis:2003px, Kinney:2005vj}, one finds $f_{\rm NL} =5/2$, from where it seems unlikely that a cancellation could happen.

In this article, we show that there is a slightly more general class of non-Gaussian consistency relations, of which Maldacena's relation is an example. This generalization emerges from a space-time reparametrization (linking short- and long-wavelength curvature perturbations) that is realized with the help of a more general symmetry. This time, the symmetry transformation involves both a time dilation and a spatial dilation. We will show that this symmetry is ap\-pro\-xi\-mate in the case of $\epsilon \ll 1$ (where $\epsilon$ is the standard first slow-roll parameter), but exact in the case of ultra slow-roll (independently of the value of $\epsilon$). In a previous work~\cite{Mooij:2015yka} a few of us have already investigated the derivation of non-Gaussian consistency relations valid for non-attractor models using symmetry arguments. The difference between the present work and ref.~\cite{Mooij:2015yka} is that the symmetry used here involves a space-time reparametrization affecting the action of curvature perturbations, whereas the symmetry explored in~\cite{Mooij:2015yka} corresponds to a symmetry of the full action driving inflation.\footnote{While completing this work we have become aware that Finelli et al. \cite{Enrico} are finishing an article on the same subject, possibly arriving to similar conclusions.}

The existence of a more general consistency relation (coming from space-time reparametrizations) suggests that the vanishing of eq.~(\ref{fNL=0}) may be effective under more general conditions, valid beyond the attractor single field models of inflation. In particular, one could expect~(\ref{fNL=0}) to be valid in the extreme case of ultra slow-roll inflation. We will argue that this is indeed the case in a companion article~\cite{us}, where the use of conformal Fermi coordinates is considered for the case of non-attractor models.

We have organized this article as follows: In Section~\ref{s2:Review-consistency} we offer a review of the derivation of the standard consistency relation for single field slow-roll inflation (attractor inflation). In Section~\ref{s3:Generalized-consistency} we derive the generalized version of the consistency relation. We do this first for the simple case $\epsilon \to 0$, and then extend this result to the more subtle case $\epsilon \neq 0$, where we pay some attention to the particular case of ultra slow-roll inflation. Then, in Section~\ref{s4:Discussion} we briefly discuss our results, and ask how they could be modified by deviations from the canonical models of inflation for which our results are strictly valid. Finally, in Section~\ref{s5:Conclusions} we present our conclusions.

\setcounter{equation}{0}
\section{Review of the consistency relation derivation}
\label{s2:Review-consistency}

Let us start by reviewing the derivation of the standard consistency relation for single field slow-roll attractor inflation, in which the curvature perturbation freezes on superhorizon scales. We will closely follow the discussion of ref.~\cite{Cheung:2007sv}, (see also the derivations in refs.~\cite{Creminelli:2004yq, Ganc:2010ff}), but with a perspective that will show to be useful for generalizing the relation later on. 

The metric line element describing a perturbed FRW spacetime, in co-moving gauge may be written as:
\be
ds^2 =  a^2 (\tau)  \bigg[ - N^2 d \tau^2  + 2  N_i d\tau dx^i + e^{2 \zeta } dx^2 \bigg] , \label{conformal-metric}
\ee
where $a$ is the usual scale factor. We have adopted conformal time $\tau$, which is related to cosmological time $t$ via $d\tau = dt / a$. The lapse $\delta N = N - 1$ and shift $N_i$ functions respect constraint equations that are found by varying the action of the perturbations. The linear solutions are given by:\footnote{In this work we assume regular Bunch-Davies initial conditions. For a discussion on the effect of considering different initial states, see \cite{Kundu:2013gha}.}
\be
\delta N = \frac{1}{\mathcal H} \partial_0 \zeta , \qquad N_i = - \partial_i \frac{\zeta}{\mathcal H} + \epsilon \frac{\partial_i}{ \partial^{2}} \partial_0 \zeta . \label{constraints-N-1}
\ee
After replacing these solutions back into the action, one obtains a cubic action describing a single scalar degree of freedom $\zeta$. Now, let us consider the following transformations of coordinates and fields:
\bea
&& x = e^{ g }  x' , \label{rep-1} \\ 
&& \tau = \tau' , \label{rep-2} \\ 
&& \zeta  = \zeta' + \Delta \zeta,  \label{rep-3}
\eea
where $g$ and $\Delta \zeta$ are functions of $\tau'$ only. We would like to know how these relations affect the form of the $\zeta$-action for a certain choice of $g$ and $\Delta \zeta$. Given that $g$ and $\Delta \zeta$ are taken as perturbations, this would require us to consider the full initial action, Einstein-Hilbert plus scalar field, including the background contributions (this is because (\ref{rep-1}) implies that some background terms will be promoted to perturbations). Instead of examining this change by inserting (\ref{rep-1})-(\ref{rep-3}) in the full action explicitly, we may analyze the way in which the metric~(\ref{conformal-metric}) is affected. This will allow us to infer how the action itself is affected by the transformation. To proceed, first notice that (\ref{rep-1}) and (\ref{rep-2}) imply
\bea
&& d x^i = e^g d x'^i + e^g \partial_0 g  x'^i d \tau' , \label{inf-1} \\
&& d \tau = d \tau' .  \label{inf-2}
\eea
In second place, recall that $N$ and $N_i$ were already fixed in terms of $\zeta$, and so they must change according to (\ref{rep-3}). This is because we are examining how the transformations alter the form of the $\zeta$-action after $N$ and $N_i$ were solved. One finds:
\bea
\delta N &=& \delta N' + \frac{1}{\mathcal H} \partial_0 \Delta \zeta , \\
N_i &=& N'_i + \partial_i \Delta \psi ,
\eea
where $\Delta \psi$ is such that
\be
\partial^2 \Delta \psi = - \partial^2 \frac{\Delta \zeta}{\mathcal H} + \epsilon  \partial_0 \Delta \zeta .
\ee
Given that we are choosing $\Delta \zeta$ to be $x'$-independent, $\Delta \psi$ satisfies the simpler equation $\partial^2 \Delta \psi =  \epsilon  \partial_0 \Delta \zeta$. This equation is solved by $\Delta \psi = \frac{1}{6} x^i x_i \epsilon  \partial_0 \Delta \zeta$, and so we may write:
\be
\partial^i \Delta \psi = \frac{1}{3} x'^i \epsilon  \partial_0 \Delta \zeta .
\ee
Then, replacing all of these results back into the metric (\ref{conformal-metric}), we obtain:
\bea
ds^2 &=& a^2 ( \tau' ) \bigg[ - e^{2 \delta N' + \frac{2}{\mathcal H} \partial_0 \Delta \zeta  } d \tau'^2    \nn \\ 
&& + 2 \Big(  N'_i + \partial_0 g  x'_i + \frac{1}{3} x'_i \epsilon  \partial_0 \Delta \zeta \Big) d\tau' dx'^i + e^{2 \zeta' + 2 \Delta \zeta  + 2 g } d x'^2 \bigg].
\eea
It is important to keep the perturbations appearing in the term proportional to $dx'^2$ up to third order at least. In this case, we have kept $\Delta \zeta$ and $g$ exactly as they appear from the definition of the transformations (\ref{rep-1})-(\ref{rep-3}). On the other hand, in those terms proportional to $d\tau'^2$ and $d\tau' dx'^i$ we must keep the perturbations up to first order at least. The reason for doing this is that we want to understand how (\ref{rep-1})-(\ref{rep-3}) change the form of the $\zeta$-action up to third order. Given that the cubic action depends on the linear contributions to $\delta N$ and $N_i$, we do not need to worry about contributions coming from $\Delta \zeta$ and $g$ beyond linear order in terms proportional to $d\tau'^2$ and $d\tau' dx'^i$.

Next, notice that if we choose both $g$ and $\Delta \zeta$ constant, and demand them to satisfy $\Delta \zeta = - g$ we end up with
\bea
ds^2 &=& a^2 ( \tau' ) \bigg[ -  N'^2 d \tau'^2  + 2 N'_i d\tau' dx'^i + e^{2 \zeta' } d x'^2 \bigg] . \label{metric-prime}
\eea
This metric has exactly the same form of (\ref{conformal-metric}), and therefore the action for $\zeta'$, obtained by using this metric, has the same form as the one for $\zeta$. This in turn, implies that both $\zeta$ and $\zeta'$ are solutions of the same system of equations of motion. Moreover, these solutions are connected through the relation:
\be
\zeta (\tau , x) = \zeta' (\tau' , x') - g .
\ee
Since $\tau = \tau'$ and $x = e^{g} x'$, we may write instead:
\be
\zeta (\tau , x) = \zeta' (\tau , e^{-g} x) - g . \label{zeta-zeta-1}
\ee
This relation may be used to derive the squeezed limit of the bispectrum in terms of the power spectrum of the perturbations. First, let us consider a splitting of $\zeta$ into short- and long-wavelength contributions of the form:
\be
\zeta = \zeta_S + \zeta_L .
\ee
\red{This separation of scales is not directly related to the size of the horizon during inflation. Initially the wavelengths of both $\zeta_L$ and $\zeta_S$ fit inside the horizon, while at later times they are both of superhorizon size. The point here is that, independent of the size of the horizon, we want to understand the non-linear effect of the long mode on the short mode.

At length scales of order $k_{S}^{-1}$, the mode $\zeta_L$ is effectively $x$-independent.} In addition, if we are interested in attractor models of single field inflation, $\zeta_L$ is also $\tau$-independent. Then, if in eq.~(\ref{zeta-zeta-1}) we choose $g = - \zeta_L$ (or, equivalently $\Delta \zeta = \zeta_L$), we end up with
\be
\zeta_S (\tau , x)  = \zeta' (\tau , e^{\zeta_L} x) . \label{short-prime-1}
\ee
In other words, the long wavelength mode of $\zeta$ has been absorbed via a coordinate transformation.\footnote{This reveals that $\zeta$ corresponds to an adiabatic mode~\cite{Weinberg:2003sw, Creminelli:2012ed}, and that the evolution of the short wavelength contribution $\zeta_S (\tau , x)$ may be thought of as that of a perturbation $\zeta'$ on a new redefined background (obtained by the absorption of $\zeta_L$).} Relation (\ref{short-prime-1}) tells us that $\zeta_S (\tau , x)$ may be expressed in terms of a fluctuation $\zeta'$ that is a solution of the same system of equations satisfied by $\zeta$, but with $e^{\zeta_L} x$ instead of $x$ in the spatial argument. In other words, we have non-linear information about how the long-wavelength mode $\zeta_L$ modulates the short wavelength mode $\zeta_S$. Next, let us consider the 2-point correlation function $\langle \zeta_S (\tau , {\bf x}) \zeta_S (\tau , {\bf y}) \rangle  \equiv \langle \zeta_S \zeta_S  \rangle (\tau , |{\bf x} -{\bf y}|)$. Equation~(\ref{short-prime-1}) tells us that
\be
\langle \zeta_S \zeta_S  \rangle (\tau , |{\bf x} -{\bf y}|) = \langle \zeta' \zeta'  \rangle (\tau , e^{\zeta_L} |{\bf x} -{\bf y}|) .
\ee
Notice that $\langle \zeta' \zeta'  \rangle (\tau , |{\bf x} -{\bf y}|) $ is nothing but the usual 2-point correlation function of the curvature perturbation in co-moving gauge (because $\zeta'$ is a solution of the full system). Expanding the previous relation in powers of $\zeta_L$, we obtain
\be
\langle \zeta_S \zeta_S  \rangle (\tau , |{\bf x} -{\bf y}|) = \langle \zeta' \zeta'  \rangle (\tau , |{\bf x} -{\bf y}|) + \zeta_L \frac{d}{d \ln |{\bf x} - {\bf y}|} \langle \zeta' \zeta'  \rangle (\tau , |{\bf x} -{\bf y}|) + \cdots.
\ee
Then, by writing the fields in Fourier space as
\be
\zeta ({\bf x}) = \frac{1}{(2 \pi)^3} \int d^3 k \zeta({\bf k}) e^{ i {\bf k} \cdot {\bf x}} ,
\ee
we end up with
\bea
 \langle  \zeta_S \zeta_S \rangle ({\bf k}_1 ,{\bf k}_2)  =  \langle \zeta' \zeta' \rangle ({\bf k}_1 ,{\bf k}_2) -  \zeta_L ({\bf k}_L)  \left[ n_s(k_S , \tau)- 1 \right]  P_\zeta (\tau, k_S) ,   \label{power-S}
\eea
where we have defined ${\bf k}_L = {\bf k}_1 + {\bf k}_2$ and ${\bf k}_S = ( {\bf k}_1 - {\bf k}_2 ) / 2$. In the previous expressions, the power spectrum $P_\zeta (\tau, k) $ and its spectral index $n_s (k) - 1$ are defined as
\bea
P_\zeta (\tau, k)  =  \int d^3 r  e^{ - i  {\bf k} \cdot  {\bf r}} \left\langle  \zeta    \zeta  \right\rangle (\tau , r) , \\
n_s(k , \tau)- 1 =  \frac{\partial }{\partial \ln k} \ln ( k^3 P_\zeta (\tau, k) ) ,
\eea
with ${\bf r} \equiv |{\bf x} -{\bf y}| $.

The first term at the rhs of eq.~(\ref{power-S}) is independent of $\zeta_L$, so by correlating eq.~(\ref{power-S}) with $\zeta_L ({\bf k}_3)$, we obtain 
\bea
\langle  \zeta_L ({\bf k}_3)  \langle  \zeta_S \zeta_S \rangle ({\bf k}_1 ,{\bf k}_2) \rangle =   - \langle  \zeta_L ({\bf k}_3)  \zeta_L ({\bf k}_L)  \rangle \left[ n_s(k_S , \tau)- 1 \right]  P_\zeta (\tau, k_S) .   \label{power-S-corr}
\eea
The squeezed limit of the bispectrum appears as the formal limit:
\be
\lim_{k_3 \to 0}(2 \pi)^3 \delta ({\bf k}_1 + {\bf k}_2 +{\bf k}_3) B_{\zeta}  ({\bf k}_1 , {\bf k}_2 , {\bf k}_3)  = \langle  \zeta_L ({\bf k}_3) \langle \zeta_S \zeta_S \rangle ({\bf k}_1 ,{\bf k}_2)  \rangle . \label{squeezed-def}
\ee
Thus, putting together eqs.~(\ref{power-S-corr}) and (\ref{squeezed-def}) we see that the squeezed limit acquires the form:
\bea
B_{\zeta}  ({\bf k}_1 , {\bf k}_2 , {\bf k}_3)   &=& - \left[ n_s(k_S , \tau)- 1 \right] P_\zeta ( k_S) P_\zeta ( k_L)  . \label{squeezed-1}
\eea
This corresponds to  Maldacena's well known consistency relation. It was obtained with the help of transformation~(\ref{short-prime-1}) linking short- and long-wavelength co-moving curvature perturbations $\zeta_S$ and $\zeta_L$ through a ``complete'' curvature perturbation $\zeta'$ (that is, a curvature perturbation for which there has been no separation of scales). In other words, (\ref{squeezed-1}) gives us information on how the long wavelength mode $\zeta_L$ modulates the short wavelength mode $\zeta_S$.

\setcounter{equation}{0}
\section{A generalized consistency relation}
\label{s3:Generalized-consistency}

We would like to count with a consistency relation valid for cases in which the long mode $\zeta_L$ is time dependent, that is, when the curvature perturbation evolves on super-horizon scales. For simplicity, let us first attempt this in the formal limit $\epsilon \to 0$. We will consider the case $\epsilon \neq 0$ in Section~\ref{case-epsilon-neq-0}. 

\subsection{Case with $\epsilon \to 0$} \label{case-epsilon-eq-0}

If $\epsilon = 0$, the Hubble parameter $H = \dot a / a$ is a constant, and the scale factor $a$ is given by
\be
a(\tau) = - \frac{1}{H \tau} ,
\ee
Then, let us consider the following transformations:
\bea
&& x = e^{ g }  x' , \label{rep-4} \\ 
&& \tau = e^{ f } \tau'  , \label{rep-5} \\ 
&& \zeta  = \zeta' + \Delta \zeta . \label{rep-6} 
\eea
Here the quantities $g$, $f$ and $\Delta \zeta$ are all functions of $\tau'$. For concreteness, let us assume that $\tau = \tau'$ at a given reference time $\tau_*$. This implies that $f = 0$ at $\tau' = \tau_*$. To make this explicit, one could write $f$ as $f (\tau' ) = \int^{\tau'}_{\tau_*} \!\! d\tau h$ (this will not be important though). This choice is completely arbitrary, and one could certainly fix initial conditions for $f$ and $g$ in other ways, without modifying the main conclusions of this section. The change of coordinates implies:
\bea
&& d x^i = e^g d x'^i + e^g \partial_0 g  x'^i d \tau' , \\
&& d \tau =e^f d \tau'(1+\tau'\partial_0 f ) .
\eea
Note that now $\partial_0 \equiv \partial_{\tau'}$. Replacing these relations back into the metric (\ref{conformal-metric}), we find:
\bea
ds^2 &=& a^2 ( \tau' ) \bigg[ - e^{2 \delta N' - 2 \tau' \partial_0 \Delta \zeta + 2 \tau' \partial_0 f  } d \tau'^2   \nn \\ 
&& + 2 \Big( N'_i + \partial_0 g  x'_i  \Big) d\tau' dx'^i + e^{2 \zeta' + 2 \Delta \zeta  + 2 g - 2 f } d x'^2 \bigg] .
\eea
As before, let us recall that the perturbations appearing together with $\delta N$ and $N_i$ may be treated up to linear order. On the other hand, those appearing together with $\zeta'$ must be treated up to cubic order. In this case, we are treating them exactly. Now, notice that if we demand that $g$ is constant, and that
\bea
  \Delta \zeta  +  g -  f  = 0 ,
\eea
the metric reduces back to (\ref{metric-prime}).  Then, we conclude that the $\zeta$-action is invariant under the transformations (\ref{rep-4})-(\ref{rep-6}). Therefore, we have two solutions $\zeta$ and $\zeta'$ related through the following relation
\be
\zeta (\tau , x) = \zeta' (e^{-f} \tau , e^{-g} x) - g + f. \label{zeta-zeta-2} 
\ee
In order to deduce the squeezed limit of the bispectrum in this class of models, let us now again consider the splitting 
\be
\zeta = \zeta_S + \zeta_L .
\ee
Recall that this time we are assuming that $\zeta_L$ depends on time. As we did with (\ref{zeta-zeta-1}), let us choose $f$ and $g$ in such a way that $\Delta \zeta = \zeta_L$:
\be
- g + f = \zeta_L (\tau) .
\ee
\red{At this point we notice that the difference between $f$ and $g$ is a pure perturbation. At the background level, where perturbations are absent, $f$ and $g$ necessarily have to coincide, and we recover the well-known De Sitter isometry $\tau \to e^\lambda \tau$, $x \to e^\lambda x$, studied in, for example, \cite{Creminelli:2012ed,Assassi:2012zq}.}

Given that $f=0$ for $\tau=\tau_*$, the previous relation sets the constant $g$ as $g= - \zeta_L (\tau_*)$. Then we find that $f$ is given by
\be
 f = \zeta_L (\tau) - \zeta_L (\tau_*) .
\ee
This leads to a relation between $\zeta_S$ and $\zeta' $ given by:
\be
\zeta_S (\tau , x)  = \zeta' ( e^{- \left[ \zeta_L (\tau) - \zeta_L (\tau_*) \right]} \tau , e^{\zeta_L (\tau_*)} x)  . \label{short-prime-2}
\ee
If $\zeta_L (\tau)$ does not evolve, then $\zeta_L (\tau) = \zeta_L (\tau_*)$, and we recover eq.~(\ref{short-prime-1}). We may now compute the power spectrum of $\zeta_S$. Up to first order in $\zeta_L$, it is direct to find in Fourier space
\bea
 \langle  \zeta_S \zeta_S \rangle ({\bf k}_1 ,{\bf k}_2)  &=&  \langle \zeta' \zeta' \rangle ({\bf k}_1 ,{\bf k}_2) -  \left[  \zeta_L ({\bf k}_L) - \zeta_L^* ({\bf k}_L )\right] \frac{d}{d \ln \tau}  P_\zeta (\tau, k_S)  \nn \\ 
 && -  \zeta_L^* ({\bf k}_L)  \left[ n_s(k_S , \tau)- 1 \right]  P_\zeta (\tau, k_S) .   \label{power-S-2}
\eea
Correlating this expression with $\zeta_L ({\bf k}_3)$, we end up with
\bea
\langle  \zeta_L ({\bf k}_3)  \langle  \zeta_S \zeta_S \rangle ({\bf k}_1 ,{\bf k}_2) \rangle &=&  - \langle  \zeta_L ({\bf k}_3)  \left[  \zeta_L ({\bf k}_L) - \zeta_L^* ({\bf k}_L )\right] \rangle  \frac{d}{d \ln \tau}   P_\zeta (\tau, k_S) \nn \\
&& - \langle  \zeta_L ({\bf k}_3)  \zeta_L^* ({\bf k}_L)  \rangle \left[ n_s(k_S , \tau)- 1 \right]  P_\zeta (\tau, k_S) .  \label{power-S-corr-2}
\eea
This expression involves the correlation of $\zeta_L ({\bf k}_3)$ evaluated at a given time $\tau$, with $\zeta_L^* ({\bf k}_3)$ which is evaluated at the reference time $\tau=\tau_*$. When superhorizon modes freeze, the first line cancels and there is no difference between  $\zeta_L^* ({\bf k}_3)$ and  $\zeta_L ({\bf k}_3)$, so we end up with Maldacena's standard attractor result.  However, if $\zeta_L$ grows on superhorizon scales fast enough for $\zeta_L^*$ to become  subdominant, and for the first line  to dominate the second one, we end up with
\bea
B_{\zeta}  ({\bf k}_1 , {\bf k}_2 , {\bf k}_3)   &=& -P_\zeta ( k_L) \frac{d}{d \ln \tau} P_\zeta ( k_S)  . \label{squeezed-2}
\eea
This is one of our main results. Equation~(\ref{squeezed-2}) tells us that under a substantial super-horizon growth \red{during inflation} the squeezed limit is dominated by a time derivative of the power spectrum.

\subsection{Non-Gaussianity in ultra slow-roll inflation}

Before considering the more general case in which $\epsilon \neq 0$, let us briefly analyze (\ref{squeezed-2}) in the context of ultra slow-roll inflation, where the inflaton field moves over a constant potential and, as a consequence, the curvature perturbation evolves exponentially after horizon crossing. The salient feature of this model is the rapid decay of $\epsilon$, which is found to be given by
\be
\epsilon \propto \frac{1}{H^2 a^6} . \label{behavior-epsilon-usr}
\ee
Although $\epsilon \to 0$ very fast, the value of $\eta$ is found to be large:
\be
\eta = -  6 \left( 1 - \frac{\epsilon}{3} \right) . \label{eta-usr}
\ee
The linear equation of motion respected by $\zeta$ on super-horizon scales is given by
\be
\frac{d}{dt} \left(  \epsilon a^3 \dot \zeta  \right) = 0 .  \label{behavior-usr}
\ee
Then, neglecting terms subleading in $\epsilon$, one finds that $\zeta  \propto \tau^{-3}$. In other words, the power spectrum on superhorizon scales behaves as:
\be
P_\zeta ( k) \propto \tau^{-6} .
\ee
Using this result back into (\ref{squeezed-2}), we find that the bispectrum in ultra slow-roll is given by
\bea
B_{\zeta}  ({\bf k}_1 , {\bf k}_2 , {\bf k}_3)   &=& 6 P_\zeta ( k_L) P_\zeta ( k_S)  , \label{squeezed-usr}
\eea
which coincides with the well known expression previously found in the literature~\cite{Namjoo:2012aa, Mooij:2015yka}. 

One should be careful with the result~(\ref{squeezed-usr}), even though it coincides with the known squeezed limit for ultra slow-roll inflation. Recall that we are judging the effect of the transformations (\ref{rep-4})-(\ref{rep-6}) on the $\zeta$-action from their effect on the metric. This implies that we are neglecting terms proportional to $\epsilon$ in the metric that could, according to eq.~(\ref{eta-usr}), have a sizable impact on the action due to time derivatives of $\epsilon$. Strictly speaking, at this point in our derivation the result of eq.~(\ref{squeezed-2}) is valid as long as $\epsilon \ll 1$ together with $\eta \ll 1$. But under these conditions it is hard (or impossible) to have a sizable super-horizon growth of $\zeta$ that could lead to an interesting situation where eq.~(\ref{squeezed-2}) could be used. To understand this issue more closely, let us analyze the case $\epsilon \neq 0$ in what follows.

\subsection{Case with $\epsilon \neq 0$} \label{case-epsilon-neq-0}

Let us now analyze the more general case in which $\epsilon \neq 0$. Here, we may consider the following transformation of coordinates and fields:
\bea
&& x = e^{ g }  x' , \label{rep-7} \\ 
&& a(\tau) = e^{- f } a(\tau') , \label{rep-8} \\ 
&& \zeta  = \zeta' + \Delta \zeta . \label{rep-9} 
\eea
Notice that we are defining the time reparametrization in terms of the scale factor $a$ in order to keep the transformation in the spatial part of the metric (which involves $a(\tau)$) valid to all orders in the perturbation $f$. The effect of this transformation on the rest of the metric may be computed up to linear order. With this in mind, it is possible to derive that the time reparametrization to linear order is given by $\tau = \tau' - \frac{1}{\mathcal H} f$, where $\mathcal H = a^{-1} \partial_0 a$. Then, the transformations lead to:
\bea
&& d x^i = e^g d x'^i + e^g \partial_0 g  x'^i d \tau' , \\
&& d \tau = d \tau' + \Big(  (1 - \epsilon ) f -  \frac{1}{\mathcal H} \partial_0 f \Big) d \tau' ,
\eea
where we used $\partial_0 \mathcal H =  (1 - \epsilon ) \mathcal H^2$. Plugging these transformations back into the action (\ref{conformal-metric}), one finds: 
\bea
ds^2 &=& a^2 ( \tau' ) \bigg[ - e^{2 \delta N' + 2 \frac{1}{\mathcal H} \partial_0 \Delta \zeta - 2 \epsilon f - 2 \frac{1}{\mathcal H} \partial_0 f  } d \tau'^2   \nn \\ 
&& + 2 \big( N'_i + \partial_0 g  x'_i  + \frac{1}{3} x'_i \epsilon  \partial_0 \Delta \zeta  \big) d\tau' dx'^i + e^{2 \zeta' + 2 \Delta \zeta  + 2 g - 2 f } d x'^2 \bigg] .
\eea
Now, consider the following conditions on $g$ and $f$:
\bea
&&   \partial_0 \Delta \zeta - \epsilon \mathcal H  f - \partial_0 f  = 0 ,  \label{Delta_zeta_0-f} \\ 
&&  \Delta \zeta  +  g -  f  =0  . \label{ref-g-f}
\eea
It is direct to see that these two equations imply:
\be
\partial_0 g =  - \epsilon \mathcal H  f . \label{partial-0-g}
\ee
Then, the metric becomes 
\bea
ds^2 &=& a^2 ( \tau' ) \bigg[ - e^{2 \delta N' } d \tau'^2  + 2 \big( N'_i + \Delta N_i   \big) d\tau' dx'^i + e^{2 \zeta' } d x'^2 \bigg] , \label{metric-pre-usr}
\eea
where we have defined $\Delta N_i$ as
\be
\Delta N_i =   - \epsilon \mathcal H  f  x'_i  + \frac{1}{3} x'_i \epsilon ( \epsilon \mathcal H  f + \partial_0 f ) ,
\ee
and where $f$ is such that it is a solution of eq.~(\ref{Delta_zeta_0-f}). Now, it is clear from this result that the $\zeta$-action will not be invariant under the present transformation unless either $\Delta N_i =0$, or $\Delta N_i$ leads to the appearance of a total derivative. This second option will not be true in general, and $\Delta N_i$ will imply terms in the action that are proportional to $\epsilon$ and $\eta$.

At this point, the metric of eq.~(\ref{metric-pre-usr}) differs from the original metric of eq.~(\ref{conformal-metric}) by the fact that $\Delta N_i$ does not vanish. The difference is of order $\epsilon$, as expected from the analysis of Section~\ref{case-epsilon-eq-0}. In what follows, let us explore what would be required to satisfy the condition $\Delta N_i =0$, independently of the size of $\epsilon$ (that is, we will not assume that $\epsilon$ is small). First, it is direct to see that $\Delta N_i =0$ is equivalent to
\be
\partial_0 ( a^{-2} \mathcal H^{-1}  f )   = 0.
\ee
This implies that $f$ must have the following dependence on time:
\be
 f    = C a^{2} \mathcal H , \label{solution-f-epsilon}
\ee
where $C$ is an integration constant that may be chosen conveniently. Note that here we cannot adopt the condition $f=0$ at a given time $\tau = \tau_*$. This is because of the way in which $f$ is introduced in eq.~(\ref{rep-8}). Now, according to eq.~(\ref{Delta_zeta_0-f}), the solution for $f$ given by eq.~(\ref{solution-f-epsilon}) must be compatible with $\Delta \zeta$. In other words, it must be possible to choose $C$ in such a way that
\be
  \partial_0 \Delta \zeta   = 3 C H^2 a^{4} , \label{Delta-zeta-C}
\ee
(where we have used $\mathcal H = H a$). This corresponds to a strong restriction on $\Delta \zeta$, which has not been chosen yet. As in the previous subsections, we are interested in identifying $\Delta \zeta$ as:
\be
\Delta \zeta = \zeta_L. \label{Delta-zeta-L}
\ee
Inserting this back into (\ref{Delta-zeta-C}), we see that $\Delta N_i =0$ is only possible if (remember that in eq.~(\ref{Delta-zeta-C}) $\d_0 \equiv \d_\tau$)
\be
\dot \zeta_L  =  3 C H^2 a^{3} . \label{zeta-L-usr}
\ee
Of course, this behavior is not guaranteed in general. However, in the particular case of ultra slow-roll inflation one has $\epsilon \propto 1 / H^2 a^6$, and so we may rewrite (\ref{zeta-L-usr}) as
\be
\dot \zeta_L  \propto  \frac{1}{\epsilon a^3}, \label{zeta-L-usr-2}
\ee
which is nothing but (\ref{behavior-usr}). As a consequence, we see that in ultra slow-roll inflation one has $\Delta N_i = 0$ independently of the value of $\epsilon$. Therefore, we have shown that the transformations (\ref{rep-7})-(\ref{rep-9}) with $f$, $g$ and $\Delta \zeta$ chosen as in (\ref{solution-f-epsilon}), (\ref{partial-0-g}), and (\ref{Delta-zeta-L}) respectively, correspond to an exact symmetry of the action for curvature perturbations in ultra slow-roll inflation (independent of the size of $\epsilon$). This should not come as a surprise. Similar to exponential inflation, ultra slow-roll inflation never ends, and so the size of $\epsilon$ (which dilutes as $\sim a^{-6}$) cannot be regarded as a fundamental quantity describing the state of inflation. 

The final step is to deduce an expression for $\zeta_S$. This is found to be
\be
\zeta_S (\tau , x)  = \zeta' ( e^{- \zeta_L  - g} \tau , e^{- g} x), \label{short-prime-3}
\ee
with $g$ the solution of eq.~(\ref{partial-0-g}). It is straightforward to see that $g$ will contribute terms that are subleading in $\epsilon$, and so we recover the expression (\ref{squeezed-2}) found in Section~\ref{case-epsilon-eq-0}. This in turn, leads to the well known result (\ref{squeezed-usr}).

\setcounter{equation}{0}
\section{Discussion}
\label{s4:Discussion}

Now that we know that (\ref{squeezed-2}) is valid for ultra slow-roll inflation, but not for general situations with $\epsilon \neq 0$, we would like to anticipate how this result could change once we consider models that depart from the exact ultra slow-roll behavior. First, if the action describing single field inflation is canonical, then all of the couplings appearing in the $\zeta$-action will consist of time derivatives of $H$, such as $\epsilon$ and $\eta$. Given that the action remains invariant under the set of transformations (\ref{rep-7})-(\ref{rep-9}) in the case of ultra slow-roll, then models with a background close to ultra slow-roll have departures at most proportional to
\be
6 + \eta . \nn 
\ee
However, in order to have a small spectral index in models close to ultra slow-roll it is necessary to have $|6 + \eta| \ll 1$, and so it would not be possible to have large departures from (\ref{squeezed-2}) unless the spectral index becomes incompatible with observations. Another possibility is to consider non-canonical models of inflation. In this class of models one has an additional parameter, the sound speed $c_s$, which is not directly related to variations of $H$. This time, the action for $\zeta$ could have terms (parametrizing departures from the ultra slow-roll case) proportional to:
\be
\left( 1 - \frac{1}{c_s^2} \right) \eta . \nn
\ee
This type of departure would not be suppressed for small values of $c_s$, and one could expect sizable modifications to the result shown in (\ref{squeezed-usr}). In fact, a direct computation shows that the modification to (\ref{squeezed-usr}) due to $c_s$ is given by~\cite{Chen:2013aj}
\bea
B_{\zeta}  ({\bf k}_1 , {\bf k}_2 , {\bf k}_3)   &\simeq& \frac{6}{c_s^2} P_\zeta ( k_L) P_\zeta ( k_S)  . \label{squeezed-usr-cs}
\eea
This result has also been obtained through symmetry arguments~\cite{Mooij:2015yka} pertaining the structure of the Lagrangian of $P(X)$-theories of inflation~\cite{Garriga:1999vw}, but not through symmetry arguments related to space-time parametrizations, as considered here. Given that $c_s$ appears as a consequence of non-gravitational interactions, it seems reasonable to assume that a space-time reparametrization leading to (\ref{squeezed-usr-cs}) does not exist.

\setcounter{equation}{0}
\section{Conclusions}
\label{s5:Conclusions}

We have generalized the well known non-Gaussian consistency relation (\ref{power-S-corr}) to a broader class of relations that is able to cope with those classes of models where the curvature perturbation $\zeta$ evolves on super-horizon scales. This relation is given by eq.~(\ref{power-S-corr-2}), and in the case where the super-horizon growth dominates, it leads to:
\bea
B_{\zeta}  ({\bf k}_1 , {\bf k}_2 , {\bf k}_3)   &=& -P_\zeta ( k_L) \frac{d}{d \ln \tau} P_\zeta ( k_S)  . \label{squeezed-conclusions}
\eea
The standard non-Gaussian consistency relation (\ref{power-S-corr}) can be understood a symmetry in\-vol\-ving a simultaneous spatial dilation and a reparametrization of the curvature perturbation. In the case of (\ref{squeezed-conclusions}), the symmetry involves a time dilation together with a reparametrization of the curvature perturbation. In both cases, the reparametrization is induced by super-horizon evolution of the long-wavelength contributions of the curvature perturbation. While this symmetry is approximate in general when $\epsilon \ll 1$, it becomes exact in the case of ultra slow-roll, independently of the value of $\epsilon$. (It is also exact when $\epsilon =0$.)

Our result complements previous studies on consistency relations derived from symmetries of quasi-de Sitter spacetimes~\cite{Creminelli:2012ed, Hinterbichler:2012nm, Senatore:2012wy, Assassi:2012zq, Hinterbichler:2013dpa, Goldberger:2013rsa, Creminelli:2013cga} applied to the context in which curvature perturbations freeze at horizon crossing. In addition, our result substantiates one more time the well known violation to the standard consistency relation found by the authors of ref.~\cite{Namjoo:2012aa}. However, our result raises the question how the non-Gaussianity expressed in (\ref{squeezed-conclusions}) would survive the transition from a non-attractor phase ---in which ultra slow-roll is dominant--- to an attractor phase where standard slow-roll inflation is dominant (before inflation ends). 

Given that the expression leading to (\ref{squeezed-conclusions}) involves a time derivative of the power spectrum, one may suspect that once the non-attractor phase concludes, and the modes stop evolving on super-horizon scales, this contribution would become suppressed. In this case, the transition to the new phase would imply a leading contribution to the bispectrum dictated by the scale dependence of the power spectrum (through $n_s - 1$). Strictly speaking, our expression cannot describe this transition. This is because during such a transition the system is no longer invariant under the set of transformations (\ref{rep-7})-(\ref{rep-9}). \red{It is invariant during ultra slow-roll, and during slow-roll, but not in between.}

One could speculate that in such a transition (from non-attractor to attractor, see also \cite{Cai:2016ngx}) the amount of non-Gaussianity in the form of (\ref{squeezed-conclusions}) could be transferred to a form of non-Gaussianity that is described by (\ref{power-S-corr}). But this would necessarily imply an unacceptably large value of the spectral index $n_s - 1$. Another possibility is that, instead of (\ref{squeezed-conclusions}), the bispectrum produced during ultra slow-roll has to be read as
\bea
B_{\zeta}  ({\bf k}_1 , {\bf k}_2 , {\bf k}_3)   &\simeq& 6 P_\zeta ( k_L) P_\zeta ( k_S)  , \label{squeezed-usr-conclusions}
\eea
without taking into consideration the time derivative appearing in the expression preceding it. In other words, the factor $6$ implied by the $\tau$-derivative becomes engraved on the distribution of superhorizon modes, and survives until the modes re-enter the horizon much after inflation.  Given that ultra slow-roll inflation has gained some prominence as a transient period of inflation that could explain certain phenomena associated to primordial physics~\cite{Germani:2017bcs, Dimopoulos:2017ged}, this seems to be a relevant issue to clarify.\footnote{\red{The recent paper \cite{Cai:2017bxr} has found that, in comoving coordinates, the transition from non-attractor to attractor inflation is characterized by a reduction of non-Gaussianity. The amount of non-Gaussianity that survives depends crucially on the nature of this transition. Roughly speaking: a sharper transition leads to more surviving non-Gaussianity.}}

 \red{Ultimately, however, we are interested in the amount of squeezed non-Gaussianity available to a free-falling observer like us, rather than to a comoving observer.} In~\cite{us} we will study this issue more closely, by introducing the use of conformal Fermi coordinates~\cite{Pajer:2013ana, Dai:2015rda, Cabass:2016cgp}. There, we will argue that the primordial squeezed non-Gaussianity produced in non-attractor models such as ultra slow-roll is non-observable ($f_{\rm NL}^{\rm obs} = 0$). \red{That would render irrelevant the question to which extent the non-attractor corrections to $f_{\rm NL}$ in comoving coordinates, as computed in this paper, survive the end of inflation, or the transition to a slow-roll phase. Whatever comoving result one gets, we will conjecture that it will be cancelled by a similar term arising from the switch from comoving to Fermi coordinates.}

\subsection*{Acknowledgements}
We would like to thank Jorge Nore\~na and Enrico Pajer for useful discussions. GAP acknowledges support from the Fondecyt Regular project number 1171811. B.P. acknowledges the CONICYT PFCHA Magister Nacional Scholarship 2016-22161360. RB also acknowledges the support from CONICYT-PCHA Doctorado Nacional  scholarship 2016-21161504. SM is funded by the Fondecyt 2015 Postdoctoral Grant 3150126.


\begin{thebibliography}{99}

\bibitem{Maldacena:2002vr} 
  J.~M.~Maldacena,
  ``Non-Gaussian features of primordial fluctuations in single field inflationary models,''
  JHEP {\bf 0305}, 013 (2003)
  [astro-ph/0210603].

\bibitem{Creminelli:2004yq} 
  P.~Creminelli and M.~Zaldarriaga,
  ``Single field consistency relation for the 3-point function,''
  JCAP {\bf 0410}, 006 (2004)
  [astro-ph/0407059].
 
\bibitem{Seery:2005wm} 
  D.~Seery and J.~E.~Lidsey,
  ``Primordial non-Gaussianities in single field inflation,''
  JCAP {\bf 0506}, 003 (2005)
  doi:10.1088/1475-7516/2005/06/003
  [astro-ph/0503692].
 
\bibitem{Chen:2006nt} 
  X.~Chen, M.~x.~Huang, S.~Kachru and G.~Shiu,
  ``Observational signatures and non-Gaussianities of general single field inflation,''
  JCAP {\bf 0701}, 002 (2007)
  [hep-th/0605045].
  
\bibitem{Cheung:2007sv} 
  C.~Cheung, A.~L.~Fitzpatrick, J.~Kaplan and L.~Senatore,
  ``On the consistency relation of the 3-point function in single field inflation,''
  JCAP {\bf 0802}, 021 (2008)
  [arXiv:0709.0295 [hep-th]].
 
\bibitem{Ganc:2010ff} 
  J.~Ganc and E.~Komatsu,
  ``A new method for calculating the primordial bispectrum in the squeezed limit,''
  JCAP {\bf 1012}, 009 (2010)
  [arXiv:1006.5457 [astro-ph.CO]].
 
\bibitem{RenauxPetel:2010ty} 
  S.~Renaux-Petel,
  ``On the squeezed limit of the bispectrum in general single field inflation,''
  JCAP {\bf 1010}, 020 (2010)
  [arXiv:1008.0260 [astro-ph.CO]].



\bibitem{Kundu:2014gxa} 
  N.~Kundu, A.~Shukla and S.~P.~Trivedi,
  ``Constraints from Conformal Symmetry on the Three Point Scalar Correlator in Inflation,''
  JHEP {\bf 1504}, 061 (2015)
  [arXiv:1410.2606 [hep-th]].

\bibitem{Kundu:2015xta} 
  N.~Kundu, A.~Shukla and S.~P.~Trivedi,
  ``Ward Identities for Scale and Special Conformal Transformations in Inflation,''
  JHEP {\bf 1601}, 046 (2016)
  [arXiv:1507.06017 [hep-th]].

\bibitem{Gong:2017yih} 
  J.~O.~Gong and M.~Yamaguchi,
  ``Correlated primordial spectra in effective theory of inflation,''
  Phys.\ Rev.\ D {\bf 95}, no. 8, 083510 (2017)
  [arXiv:1701.05875 [astro-ph.CO]].
  
\bibitem{Tanaka:2011aj} 
  T.~Tanaka and Y.~Urakawa,
  ``Dominance of gauge artifact in the consistency relation for the primordial bispectrum,''
  JCAP {\bf 1105}, 014 (2011)
  [arXiv:1103.1251 [astro-ph.CO]].
  
\bibitem{Pajer:2013ana} 
  E.~Pajer, F.~Schmidt and M.~Zaldarriaga,
  ``The Observed Squeezed Limit of Cosmological Three-Point Functions,''
  Phys.\ Rev.\ D {\bf 88}, no. 8, 083502 (2013)
  [arXiv:1305.0824 [astro-ph.CO]].

\bibitem{Dai:2015rda} 
  L.~Dai, E.~Pajer and F.~Schmidt,
  ``Conformal Fermi Coordinates,''
  JCAP {\bf 1511}, no. 11, 043 (2015)
  [arXiv:1502.02011 [gr-qc]].

\bibitem{Cabass:2016cgp} 
  G.~Cabass, E.~Pajer and F.~Schmidt,
  ``How Gaussian can our Universe be?,''
  JCAP {\bf 1701}, no. 01, 003 (2017)
  [arXiv:1612.00033 [hep-th]].

\bibitem{Tada:2016pmk} 
  Y.~Tada and V.~Vennin,
  ``Squeezed bispectrum in the $\delta N$ formalism: local observer effect in field space,''
  JCAP {\bf 1702}, no. 02, 021 (2017)
  [arXiv:1609.08876 [astro-ph.CO]].
  
\bibitem{Yoo:2009au} 
  J.~Yoo, A.~L.~Fitzpatrick and M.~Zaldarriaga,
  ``A New Perspective on Galaxy Clustering as a Cosmological Probe: General Relativistic Effects,''
  Phys.\ Rev.\ D {\bf 80}, 083514 (2009)
  doi:10.1103/PhysRevD.80.083514
  [arXiv:0907.0707 [astro-ph.CO]].

\bibitem{Baldauf:2011bh} 
  T.~Baldauf, U.~Seljak, L.~Senatore and M.~Zaldarriaga,
  ``Galaxy Bias and non-Linear Structure Formation in General Relativity,''
  JCAP {\bf 1110}, 031 (2011)
  doi:10.1088/1475-7516/2011/10/031
  [arXiv:1106.5507 [astro-ph.CO]].

\bibitem{Manasse:1963zz} 
  F.~K.~Manasse and C.~W.~Misner,
  ``Fermi Normal Coordinates and Some Basic Concepts in Differential Geometry,''
  J.\ Math.\ Phys.\  {\bf 4}, 735 (1963).
  doi:10.1063/1.1724316
  
\bibitem{Guth:1980zm} 
  A.~H.~Guth,
  ``The Inflationary Universe: A Possible Solution to the Horizon and Flatness Problems,''
  Phys.\ Rev.\ D {\bf 23}, 347 (1981).
  
\bibitem{Linde:1981mu} 
  A.~D.~Linde,
  ``A New Inflationary Universe Scenario: A Possible Solution of the Horizon, Flatness, Homogeneity, Isotropy and Primordial Monopole Problems,''
  Phys.\ Lett.\ B {\bf 108}, 389 (1982).
  
\bibitem{Starobinsky:1980te} 
  A.~A.~Starobinsky,
  ``A New Type of Isotropic Cosmological Models Without Singularity,''
  Phys.\ Lett.\ B {\bf 91}, 99 (1980).

\bibitem{Albrecht:1982wi} 
  A.~Albrecht and P.~J.~Steinhardt,
  ``Cosmology for Grand Unified Theories with Radiatively Induced Symmetry Breaking,''
  Phys.\ Rev.\ Lett.\  {\bf 48}, 1220 (1982).
  
\bibitem{Mukhanov:1981xt} 
  V.~F.~Mukhanov and G.~V.~Chibisov,
  ``Quantum Fluctuation and Nonsingular Universe. (In Russian),''
  JETP Lett.\  {\bf 33}, 532 (1981)
  [Pisma Zh.\ Eksp.\ Teor.\ Fiz.\  {\bf 33}, 549 (1981)].
  
  
\bibitem{Sasaki:2006kq} 
  M.~Sasaki, J.~Valiviita and D.~Wands,
  ``Non-Gaussianity of the primordial perturbation in the curvaton model,''
  Phys.\ Rev.\ D {\bf 74}, 103003 (2006)
  [astro-ph/0607627].
  
\bibitem{Byrnes:2008wi} 
  C.~T.~Byrnes, K.~Y.~Choi and L.~M.~H.~Hall,
  ``Conditions for large non-Gaussianity in two-field slow-roll inflation,''
  JCAP {\bf 0810}, 008 (2008)
  [arXiv:0807.1101 [astro-ph]].
  
\bibitem{Kinney:2005vj} 
  W.~H.~Kinney,
  ``Horizon crossing and inflation with large eta,''
  Phys.\ Rev.\ D {\bf 72}, 023515 (2005)
  [gr-qc/0503017].


  
\bibitem{Namjoo:2012aa} 
  M.~H.~Namjoo, H.~Firouzjahi and M.~Sasaki,
  ``Violation of non-Gaussianity consistency relation in a single field inflationary model,''
  Europhys.\ Lett.\  {\bf 101}, 39001 (2013)
  [arXiv:1210.3692 [astro-ph.CO]].

\bibitem{Martin:2012pe} 
  J.~Martin, H.~Motohashi and T.~Suyama,
  ``Ultra Slow-Roll Inflation and the non-Gaussianity Consistency Relation,''
  Phys.\ Rev.\ D {\bf 87}, no. 2, 023514 (2013)
  [arXiv:1211.0083 [astro-ph.CO]].
  
\bibitem{Chen:2013aj} 
  X.~Chen, H.~Firouzjahi, M.~H.~Namjoo and M.~Sasaki,
  ``A Single Field Inflation Model with Large Local Non-Gaussianity,''
  Europhys.\ Lett.\  {\bf 102}, 59001 (2013)
  [arXiv:1301.5699 [hep-th]].
  
\bibitem{Mooij:2015yka} 
  S.~Mooij and G.~A.~Palma,
  ``Consistently violating the non-Gaussian consistency relation,''
  JCAP {\bf 1511}, no. 11, 025 (2015)
  [arXiv:1502.03458 [astro-ph.CO]].
  
\bibitem{Tsamis:2003px} 
  N.~C.~Tsamis and R.~P.~Woodard,
  ``Improved estimates of cosmological perturbations,''
  Phys.\ Rev.\ D {\bf 69}, 084005 (2004)
  [astro-ph/0307463].

\bibitem{Enrico}
B.~Finelli, G.~Goon, E.~Pajer and L.~Santoni,
``Soft Theorems For Shift-Symmetric Cosmologies,''
[arXiv:1711.03737 [hep-th]]

\bibitem{us} 
  R.~Bravo, S.~Mooij, G.~A.~Palma and B.~Pradenas
  ``Vanishing of local non-Gaussianity in canonical single field inflation,''
[arXiv:1711.05290 [astro-ph.CO]]

\bibitem{Kundu:2013gha} 
  S.~Kundu,
  ``Non-Gaussianity Consistency Relations, Initial States and Back-reaction,''
  JCAP {\bf 1404}, 016 (2014)
  [arXiv:1311.1575 [astro-ph.CO]].
  
\bibitem{Weinberg:2003sw} 
  S.~Weinberg,
  ``Adiabatic modes in cosmology,''
  Phys.\ Rev.\ D {\bf 67}, 123504 (2003)
  [astro-ph/0302326].
  
\bibitem{Creminelli:2012ed} 
  P.~Creminelli, J.~Nore\~na and M.~Simonovi\'c,
  ``Conformal consistency relations for single-field inflation,''
  JCAP {\bf 1207}, 052 (2012)
  [arXiv:1203.4595 [hep-th]].
  
\bibitem{Assassi:2012zq} 
  V.~Assassi, D.~Baumann and D.~Green,
  ``On Soft Limits of Inflationary Correlation Functions,''
  JCAP {\bf 1211}, 047 (2012)
  [arXiv:1204.4207 [hep-th]].


\bibitem{Garriga:1999vw} 
  J.~Garriga and V.~F.~Mukhanov,
  ``Perturbations in k-inflation,''
  Phys.\ Lett.\ B {\bf 458}, 219 (1999)
  doi:10.1016/S0370-2693(99)00602-4
  [hep-th/9904176].

\bibitem{Hinterbichler:2012nm} 
  K.~Hinterbichler, L.~Hui and J.~Khoury,
  ``Conformal Symmetries of Adiabatic Modes in Cosmology,''
  JCAP {\bf 1208}, 017 (2012)
  [arXiv:1203.6351 [hep-th]].

\bibitem{Senatore:2012wy} 
  L.~Senatore and M.~Zaldarriaga,
  ``A Note on the Consistency Condition of Primordial Fluctuations,''
  JCAP {\bf 1208}, 001 (2012)
  [arXiv:1203.6884 [astro-ph.CO]].


\bibitem{Hinterbichler:2013dpa} 
  K.~Hinterbichler, L.~Hui and J.~Khoury,
  ``An Infinite Set of Ward Identities for Adiabatic Modes in Cosmology,''
  JCAP {\bf 1401}, 039 (2014)
  [arXiv:1304.5527 [hep-th]].

\bibitem{Goldberger:2013rsa} 
  W.~D.~Goldberger, L.~Hui and A.~Nicolis,
  ``One-particle-irreducible consistency relations for cosmological perturbations,''
  Phys.\ Rev.\ D {\bf 87}, no. 10, 103520 (2013)
  [arXiv:1303.1193 [hep-th]].
  
\bibitem{Creminelli:2013cga} 
  P.~Creminelli, A.~Perko, L.~Senatore, M.~Simonovi\'c and G.~Trevisan,
  ``The Physical Squeezed Limit: Consistency Relations at Order $q^2$,''
  JCAP {\bf 1311}, 015 (2013)
  [arXiv:1307.0503 [astro-ph.CO]].

\bibitem{Cai:2016ngx} 
  Y.~F.~Cai, J.~O.~Gong, D.~G.~Wang and Z.~Wang,
  ``Features from the non-attractor beginning of inflation,''
  JCAP {\bf 1610}, no. 10, 017 (2016)
  [arXiv:1607.07872 [astro-ph.CO]].


\bibitem{Germani:2017bcs} 
  C.~Germani and T.~Prokopec,
  ``On primordial black holes from an inflection point,''
  Phys.\ Dark Univ.\  {\bf 18}, 6 (2017)
  [arXiv:1706.04226 [astro-ph.CO]].


\bibitem{Dimopoulos:2017ged} 
  K.~Dimopoulos,
  ``Ultra slow-roll inflation demystified,''
  Phys.\ Lett.\ B {\bf 775}, 262 (2017)
  [arXiv:1707.05644 [hep-ph]].

\bibitem{Cai:2017bxr} 
  Y.~F.~Cai, X.~Chen, M.~H.~Namjoo, M.~Sasaki, D.~G.~Wang and Z.~Wang,
  ``Revisiting non-Gaussianity from non-attractor inflation models,''
  arXiv:1712.09998 [astro-ph.CO].


  
  
\end{thebibliography}
\end{document}